\documentclass[]{spie}  

\usepackage{amsmath,amsfonts,amssymb}
\usepackage{graphicx}
\usepackage[colorlinks=true, allcolors=blue]{hyperref}
\usepackage{bm}
\usepackage[table,xcdraw]{xcolor}

\title{WIN-PDQ: A Wiener-estimator-based projection-domain quantitative SPECT method that accounts for intra-regional uptake heterogeneity}

\author[1]{Zekun Li}
\author[2]{Nadia Benabdallah}
\author[2]{Daniel L. J. Thorek}
\author[1,2]{Abhinav K. Jha}
\affil[1]{Department of Biomedical Engineering, Washington University in St. Louis, St. Louis, MO, USA}
\affil[2]{Mallinckrodt Institute of Radiology, Washington University in St. Louis, St. Louis, MO, USA}

\begin{document} 
\maketitle

\begin{abstract}

Single photon emission computed tomography (SPECT) can enable the quantification of activity uptake in lesions and at-risk organs in $\alpha$-particle-emitting radiopharmaceutical therapies ($\alpha$-RPTs). However, this quantification is challenged by the extremely low detected photon counts, complicated isotope physics, and the image-degrading effects in $\alpha$-RPT SPECT. 
Thus, strategies to optimize the SPECT system and protocol designs for the task of regional uptake quantification are much needed.
Objectively performing this task-based optimization requires a reliable (accurate and precise) regional uptake quantification method.
Conventional reconstruction-based quantification (RBQ) methods have been observed to be erroneous for $\alpha$-RPT SPECT. Projection-domain quantification methods, which estimate regional uptake directly from SPECT projections, have demonstrated potential in providing reliable regional uptake estimates, but these methods assume constant uptake within the regions, an assumption that may not hold. 
To address these challenges, we propose Wiener INtegration Projection-Domain Quantification (WIN-PDQ), a Wiener-estimator-based projection-domain quantitative SPECT method. The method accounts for the heterogeneity within the regions of interest while estimating mean uptake. An early-stage evaluation of the method was conducted using 3D Monte Carlo-simulated SPECT of anthropomorphic phantoms with $^{223}$Ra uptake and lumpy-model-based intra-regional uptake heterogeneity. 
In this evaluation with phantoms of varying mean regional uptake and intra-regional uptake heterogeneity, the WIN-PDQ method yielded ensemble unbiased estimates and significantly outperformed both reconstruction-based and previously proposed projection-domain quantification methods in terms of normalized root ensemble mean squared error. 
%
In conclusion, based on these preliminary findings, the proposed WIN-PDQ method is showing potential for estimating mean regional uptake in $\alpha$-RPTs and towards enabling the objective task-based optimization of SPECT system and protocol designs.

\end{abstract}


\section{Introduction}
\label{sec:intro}

Alpha-particle-emitting radiopharmaceutical therapies ($\alpha$-RPTs) have become increasingly important as a cancer treatment modality~\cite{kluetz2014radium,mcdevitt2018feed,tafreshi2019development}. Characterized by their short emission range and high linear energy transfer, $\alpha$-particles can effectively ablate the regions where they are deposited, with minimal damage to adjacent tissues~\cite{tafreshi2019development}. However, the systemic administration of these radiopharmaceuticals results in their dispersion throughout the patient body, leading to unknown levels of accumulation at sites of diseases and within radiosensitive critical organs. Thus, it is important to quantify the absorbed doses of the lesions and different organs of the patient treated with these potent agents. Such dose quantification can help with adapting treatment regimens, predicting therapy outcomes, and monitoring adverse events~\cite{brans2007clinical,garin2021personalised}.

Often, $\alpha$-emitting isotopes also emit X- and $\gamma$-ray photons, which are detectable by a \(\gamma\)-camera. Thus, single-photon emission computed tomography (SPECT) could provide a mechanism for quantifying activity uptake in organs and lesions of patients undergoing $\alpha$-RPTs. The quantified regional activity uptake can subsequently serve as input to dosimetry techniques to perform organ and lesion-level dosimetry, a key reason for conducting quantitative SPECT in $\alpha$-RPTs~\cite{sgouros2020dosimetry,pandit2021dosimetry}. However, the SPECT-based regional activity uptake quantification task is challenging in $\alpha$-RPTs. In addition to the complicated image-degrading effects, the primary challenge is the extremely low number of photon count detections with conventional SPECT system and protocol designs. This is due to the up to 1000 times lower administered activity in $\alpha$-RPTs than conventional radionuclide therapies. 
To enable reliable quantification given such very low administered activity, it is important to develop strategies to optimize the SPECT system and protocol designs based on the task of regional uptake quantification in $\alpha$-RPTs. To objectively perform such task-based optimization, a reliable (accurate and precise) quantitative SPECT method to estimate the regional uptake is needed.  

Conventional quantitative SPECT methods are based on reconstruction-based quantification (RBQ). This approach involves obtaining reconstructed SPECT images, followed by defining volumes of interest (VOIs) corresponding to different lesions and organs. Regional mean activity uptake is then calculated by averaging voxel values within each VOI. 
However, the RBQ methods were observed to be not optimal given the low counts and complicated SPECT physics in \(\alpha\)-RPTs, yielding considerable bias and variability in uptake estimates~\cite{benabdallah2021practical,benabdallah2019223,gustafsson2020feasibility,yue2016f,li2022projection}. The major challenge in RBQ methods is that the regional uptake estimation is performed from reconstructed SPECT images. Reconstructing the images requires estimating a large number of voxel values from the projection data, an inherently ill-posed problem that becomes even more challenging when the number of detected counts is small.
Further, the image reconstruction process is also subject to information loss~\cite{barrett2013foundations,beaudry2011intuitive}.
In this context, we recognize that for the task of estimating the regional uptake, voxel-based image reconstruction is an intermediary step, and notably, the number of VOIs requiring uptake estimation is significantly less than the total number of voxels. Thus, direct quantification of mean uptake in VOIs from projection data is a less ill-posed problem and it also could avoid the information loss arising in the reconstruction process.

Given the above considerations, projection-domain quantitative SPECT methods, which estimate the uptake values in lesions and different organs directly from SPECT projections—including scanning-linear~\cite{whitaker2008estimating,konik2015comparison}, Wiener~\cite{barrett2013foundations,barrett2008adaptive,lin2018task}, and maximum-likelihood (ML) estimator-based approaches~\cite{carson1986maximum,li2022projection,li2023joint,li2021Fully3D}—have shown potential in prior studies. Notably, a low-count quantitative SPECT (LC-QSPECT) approach using ML estimation~\cite{li2022projection,li2023joint,li2021Fully3D} has been developed for direct activity uptake estimation in VOIs from \(\alpha\)-RPT SPECT projections. This method can effectively compensate for the complicated SPECT physics in \(\alpha\)-RPTs and address the stray-radiation-related noise that becomes significant at low counts. The method has been observed to deliver precise and accurate regional activity uptake estimates in \(\alpha\)-RPTs, given VOI definitions with uniform intra-regional uptake~\cite{li2022projection,li2024ISITQA}.
However, existing projection-domain quantification methods assume constant uptake within the regions when estimating the mean uptake and thus are not able to account for potential intra-regional activity uptake heterogeneity.
Previous studies~\cite{phelps2021sodium,benabdallah2024beyond} have demonstrated the presence of considerable heterogeneous activity uptake within various regions of patients treated with \(\alpha\)-RPT, especially in lesions. 
Further, studies have shown that neglecting significant intra-regional uptake heterogeneity can lead to unreliable estimates of the mean regional uptake~\cite{li2023joint}.  
Therefore, there is an important need for a method to estimate mean regional uptake from the SPECT projections while effectively accounting for intra-regional uptake heterogeneity. 
Once developed, this method will provide an important tool to objectively optimize SPECT system and protocol designs for the task of mean regional uptake estimation, even in the presence of intra-regional uptake heterogeneity. 

Towards addressing this need, we propose and evaluate a Wiener-estimator-based projection-domain quantitative SPECT method, named Wiener INtegration Projection-Domain Quantification (WIN-PDQ), which estimates mean regional uptake directly from SPECT projections while accounting for the intra-regional uptake heterogeneity.


\section{Method}
\label{sec:method}

\subsection{Theory}
\label{sec:theory}
\label{sec:general_theory}
Let $f( \bm{r|\bm{\theta}} )$ be the activity uptake distribution within a patient, where $\bm{r} = (x,y,z)$ denotes the 3D spatial coordinates, and $\bm{\theta}$ represents parameters describing intra-regional uptake heterogeneity. For a SPECT system imaging the patient, denote $\bm{g}$ as the measured $M$-dimensional projection data, $\mathcal{H}$ the SPECT system operator, and $\bm{\Psi}$ an $M$-dimensional vector representing stray-radiation-related noise that becomes significant at low counts. 
We assume that the mean value of this noise is uniformly distributed across all the projection bins within a specific energy window. 
Thus, each element of $\bm{\Psi}$ equals $\psi$, the mean value of this noise in each projection bin. Then, $\bm{g}$ is Poisson distributed with mean $\mathcal{H}\bm{f} + \bm{\Psi}$, and the imaging system equation is
\begin{equation}
\label{eq:image_eq}
    \bm{g} = \mathcal{H}\bm{f} + \bm{\Psi} + \bm{n},
\end{equation}
where $\bm{n}$ represents the Poisson noise.

Denote $\lambda_{k}$ and $\phi_{k}( \bm{r} )$ as the mean regional uptake value and VOI function for the $k^{th}$ VOI in a patient with $K$ total VOIs, respectively. Similar to the conventional Wiener estimator, the WIN-PDQ method requires prior knowledge of the first and second-order statistics of the $\bm{\lambda}$ and $\bm{g}$. 
Denote the mean of each element in $\bm{\lambda}$ and the average of the projection data over all randomness, including those caused by the intra-regional heterogeneity, as $\overline{\bm{\lambda}} = < \bm{\lambda} >_{\bm{\lambda}}$ and $\overline{\overline{\overline{\bm{g}}}} = < < < \bm{g} >_{\bm{g|\lambda,\theta}} >_{\bm{\theta|\lambda}} >_{\bm{\lambda}}$, respectively, where $< \bullet >_{\bm{x}}$ denotes $\int d\bm{x} \bullet \text{pr}( \bm{x} )$ when $\bm{x}$ is continuous or $\sum_{\bm{x}}^{}{\bullet \Pr(\bm{x})}$ when $\bm{x}$ is discrete.
The general form of a globally unbiased linear estimator, characterized by a $M \times K$ dimensional matrix $\bm{W}$, is given by:
\begin{equation}
\label{eq:template}
\widehat{\bm{\lambda}}( \bm{g} ) = \overline{\bm{\lambda}} + \bm{W}^{t}( \bm{g} - \overline{\overline{\overline{\bm{g}}}} ).
\end{equation}
For estimators of this form, the ensemble mean squared error (EMSE) of the estimates, considering all randomness, is given by:
\begin{equation}
\label{eq:EMSE_ana}
\begin{aligned}
EMSE &= \text{tr}(\bm{W}^{t}\bm{K}_{\bm{g}}\bm{W}) - 2\text{tr}(\bm{K}_{\bm{\lambda g}}\bm{W}) + \text{tr}(\bm{K}_{\bm{\lambda}}),
\end{aligned}
\end{equation}
where $\text{tr}(\bm{K})$ is the trace of a matrix $\bm{K}$; $\bm{K}_{\bm{g}}$, $\bm{K}_{\bm{\lambda g}}$, and $\bm{K}_{\bm{\lambda}}$ are the covariance of $\bm{g}$, cross covariance of $\bm{\lambda}$ and $\bm{g}$, and covariance of $\bm{\lambda}$, respectively. We note here that these covariance matrices consider all sources of randomness including the randomness due to intra-regional heterogeneity. 
Obtaining the matrix $\bm{W}$ that minimizes this EMSE value gives the estimator matrix of the WIN-PDQ method.
The analytical solution of $\bm{W}$ is given by differentiating Eq.~\eqref{eq:EMSE_ana} with respect to $\bm{W}$ and settiing the derivative to zero, yielding:
\begin{equation}
\label{eq:invers_solution}
\bm{W} = \bm{K}_{\bm{g}}^{-1}\bm{K}_{\bm{\lambda g}}^{t}.
\end{equation}

Given the high dimensionality of $\bm{g}$, statistically obtaining $\bm{K}_{\bm{g}}$ and $\bm{K}_{\bm{\lambda g}}$ presents significant challenges. Consequently, further derivations to simplify these covariance matrices are needed. To obtain more analytically tractable expressions for computing these matrices, we describe the activity uptake distribution of the patient by
\begin{equation}
\label{eq:f_HetRepresentation}
f(\bm{r} | \bm{\theta}) = \sum_{k=1}^{K} \lambda_{k}\varphi_{k}(\bm{r} | \bm{\theta}_{k}),
\end{equation}
where $\varphi_{k}( \bm{r}|\bm{\theta}_{k} )$ is the normalized heterogeneous uptake distribution within the $k^{th}$ region, parameterized by $\bm{\theta}_{k}$. 
We assume that $\bm{\theta}$ is independent of $\bm{\lambda}$.
Denote $\overline{\overline{\bm{g}}} = < < \bm{g} >_{\bm{g|\lambda,\theta}} >_{\bm{\theta}}$, then:
\begin{equation}
\overline{\overline{\bm{g}}} = \ \sum_{k = 1}^{K}\lambda_{k}\bm{H}_{k} + \bm{\Psi},
\end{equation}
where $\bm{H}_{k}\mathcal{= H <}\varphi_{k}( \bm{r}|\bm{\theta}_{k} ) >_{\bm{\theta}}$ can be interpreted as the response of the system to the averaged intra-regional heterogeneity in the $k^{th}$ region. Denote the matrix with $\bm{H}_{k}$ as the $k^{th}$ column by $\bm{H}$. We can derive $\bm{K}_{\bm{\lambda g}}$ as:
\begin{equation}
\label{eq:K_lambda_g}
\bm{K}_{\bm{\lambda g}} = \bm{K}_{\bm{\lambda}}\bm{H}^{t}.
\end{equation}
Next, we can express $\bm{K}_{\bm{g}}$ as:
\begin{equation}   
\label{eq:k_g}
\bm{K}_{\bm{g}} = \bm{K}_{\bm{n}} + \overline{\bm{K}_{\overline{\bm{g}}}} + \bm{H}\bm{K}_{\bm{\lambda}}\bm{H}^{t},
\end{equation}
Given that the data $\bm{g}$ follows a Poisson distribution with independent elements, $\bm{K_n}$ is a diagonal matrix with diagonal elements from the vector $\bm{H} \overline{\bm{\lambda}} + \bm{\Psi}$.
Denote $\overline{\bm{g}} = <\bm{g}>_{\bm{g|\lambda,\theta}}$. Denote $\bm{\varphi}$ as a $K$-dimensional vector comprised of function elements. Specifically, the $k^{th}$ element of $\bm{\varphi}$, denoted as $\varphi_k$, is $\varphi_{k}(\bm{r} | \bm{\theta}_{k})$.
Derived from Eqs.~\eqref{eq:image_eq} and~\eqref{eq:f_HetRepresentation}, and making use of the linearity of the system operator, we obtain:
\begin{equation}
\label{eq:k_g_bar}
\overline{\bm{K}_{\overline{\bm{g}}}} 
= < < \lbrack \overline{\bm{g}} - \overline{\overline{\bm{g}}} \rbrack \lbrack \overline{\bm{g}} - \overline{\overline{\bm{g}}} \rbrack^{t} >_{\bm{\theta}} >_{\bm{\lambda}} 
=  < \bm{\lambda}^{t} <\lbrack \mathcal{H}\bm{\varphi} - \mathcal{H}\overline{\bm{\varphi}} \rbrack \lbrack \mathcal{H}\bm{\varphi} - \mathcal{H}\overline{\bm{\varphi}} \rbrack^{t} >_{\bm{\theta}} \bm{\lambda} >_{\bm{\lambda}},
\end{equation}
where $\mathcal{H}\bm{\varphi}$ is a $K$-dimensional vector. 
The elements of $\mathcal{H}\bm{\varphi}$ are vectors, with the $k^{th}$ element given by $ \mathcal{H} \varphi_{k}$. Additionally, $\overline{\bm{\varphi}}=<\bm{\varphi}>_{\bm{\theta}}$. 
With this theoretical framework, 
given prior knowledge of $\varphi_{k}( \bm{r}|\bm{\theta}_{k} )$ for any $\bm{\theta}_k$ and $k$, in addition to the distribution of $\bm{\theta}$,
we could solve the estimator matrix $\bm{W}$.
We also note that under the homogeneous intra-regional uptake distribution model, $\overline{\bm{K}_{\overline{\bm{g}}}}$ becomes a zero matrix, then, the proposed method reduces to the conventional Wiener estimator.

\subsection{Experiments}

\subsubsection{Phantom design}
\begin{figure}
    \centering
    \includegraphics[width=0.85\textwidth]{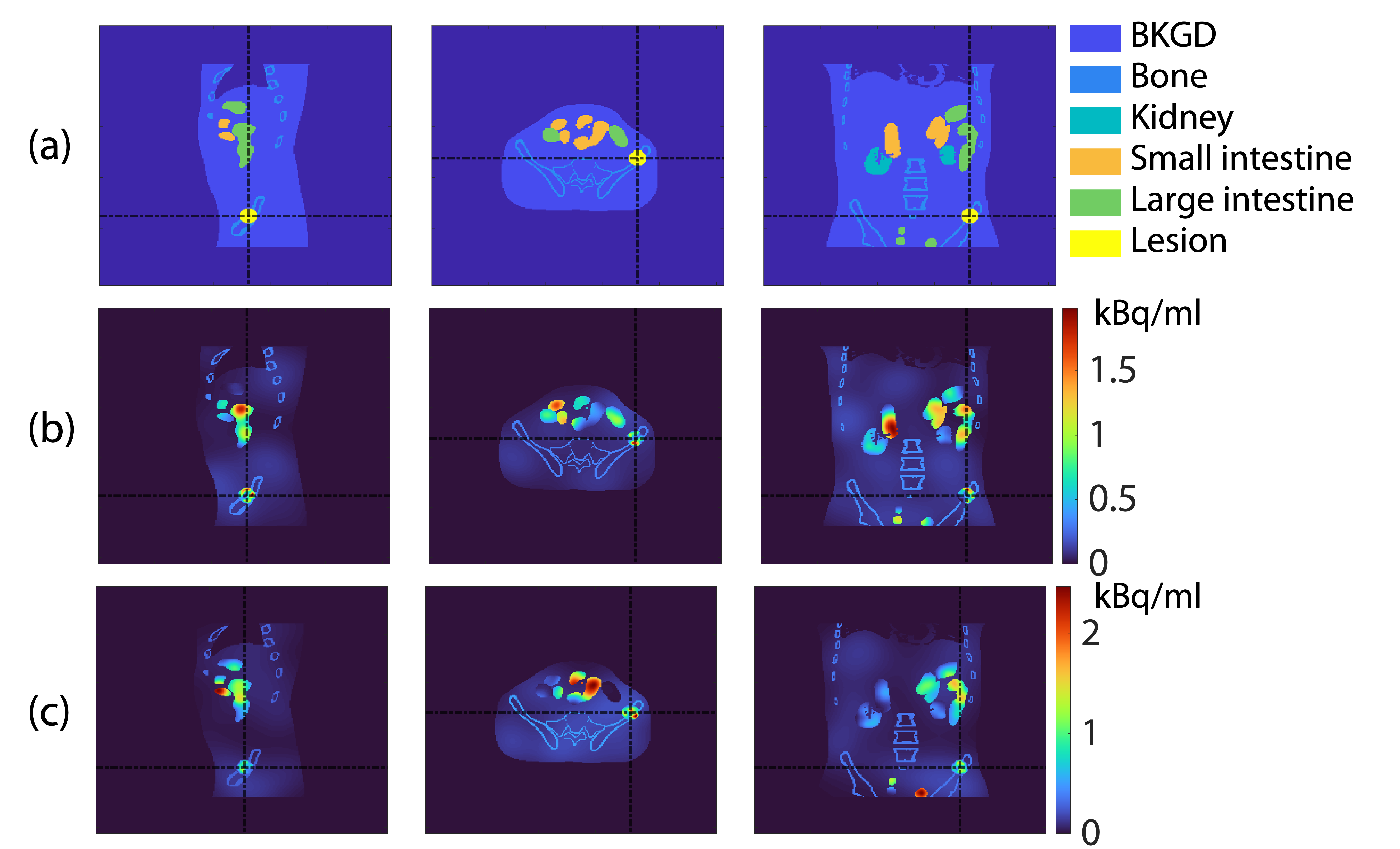}
    \caption{(a) region maps of the standard patient model. (b) and (c) show activity maps of two representative phantom realizations with heterogeneous intra-regional uptake distributions. The dashed lines indicate the relative positions of the other two planes.}
    \label{fig:phantom_img}
\end{figure}

We evaluated the WIN-PDQ method using 3D anthropomorphic phantoms for patients with bone metastatic castrate-resistant prostate cancer (bmCRPC) treated with $^{223}$Ra-based $\alpha$-RPT, the first $\alpha$-RPT approved by the U.S. FDA~\cite{fdaAppRa}. 
All phantoms in this study were generated from a standard patient model, but with varying mean regional uptake and intra-regional uptake heterogeneity. 
The attenuation and region maps of the standard patient model were created using the XCAT software. Six regions of the patient were considered, including the bone (cortical bone only), kidney, large intestine, small intestine, and a vertebral lesion, along with an additional background region encompassing all other low uptake regions in the patient~\cite{XCATpht}.
Both the region and attenuation maps had dimensions of \(256 \times 256 \times 180\), with a voxel side length of 2.209 mm.
The torso region of a patient with middle body size (representative of the 50th percentile male in the U.S.) and a lesion with a diameter of 33.75 mm, reflecting the average lesion size reported in a clinical study~\cite{murray2017potential}, was considered. 
The region maps of such a patient model are shown in Fig.~\ref{fig:phantom_img} (a).

We modeled heterogeneous activity uptake distributions within the lesion, kidney, large intestine, small intestine, and background regions.
The heterogeneity was simulated using a lumpy model with Gaussian lump functions.
More specifically, the lumpy model generates tissue-like objects by placing a random number of lumps with arbitrary positions within a VOI, creating spatial variations in the uptake. Each lump is represented by a function $\Lambda( \bullet )$, typically Gaussian. In this study, we use a normalized lumpy model, with Gaussian lump functions:
\begin{equation}
\label{eq:lumpy_obj}
\varphi_{k}(\bm{r} | \bm{\theta}_{k}) = \phi_{k}(\bm{r})\frac{A_{k}}{N_{k} + 1}\sum_{n=1}^{N_{k}+1}\frac{\Lambda(\bm{r} - \bm{c}_{n}^{k} | s_{k})}{\int d\bm{r'} \Lambda(\bm{r'} - \bm{c}_{n}^{k} | s_{k})\phi_{k}(\bm{r'})},
\end{equation}
where $A_{k} = \int d^{3}r\ \phi_{k}( \bm{r} )$ and $\bm{\theta}_{k} = \{ N_{k},\bm{c}^{k},s_{k}\}$. In this model, $N_{k}$ represents the number of lumps, $\bm{c}^{k}$ the lump center locations, and $s_{k}$ the lump size (standard deviation of the Gaussian lump function) in the $k^{th}$ VOI. We model $N_{k}$ as Poisson-distributed with mean $\overline{N_{k}}$, and $\bm{c}_{n}^{k}$ as uniformly distributed within the $k^{th}$ VOI. Both $N_k$ and $s_k$ are constant for a specific type of region.

To generate multiple phantom realizations based on the standard patient model, we first generated 10 realizations of these heterogeneous intra-regional uptake distributions.
The standard deviation of the lumps, $s_{k}$, was set to 35 mm for the background, 9 mm for the kidney, 13 mm for the large intestine, 14 mm for the small intestine, and 3 mm for the lesion. The mean number of lumps, $\overline{N_k}$, was set as 100 for all regions. Then, from each heterogeneous intra-regional uptake distribution realization, 50 additional phantom realizations were generated, each with different sampled mean regional uptake values. 
The regional uptake values were independently sampled from Gaussian distributions, with parameters specified in Table~\ref{tab:region_uptake}, to simulate a clinically realistic count level in $\alpha$-RPTs. 
Activity maps of two representative phantom realizations are presented in Fig.~\ref{fig:phantom_img} (b) and (c).

\begin{table}[]
\centering
\caption{The mean and standard deviation of the regional uptake across the phantom realizations.}
\label{tab:region_uptake}
\vspace*{2mm}
\begin{tabular}{ccccccc}
\hline
\rowcolor[HTML]{EFEFEF} 
Unit: kBq/ml & Background & Bone & Kidney & Large intestine & Small intestine & Lesion \\ \hline
Mean         & 0.07   & 0.35   & 0.35     & 1 & 1 & 0.7     \\
Std. dev.    & 0.02    & 0.1   & 0.1     & 0.3 & 0.3 & 0.2     \\ \hline
\end{tabular}
\end{table}

\subsubsection{Modeling the SPECT system}
\label{sec:SPECTsys}
Projection data corresponding to each phantom realization was obtained using a SIMIND-simulated dual-head GE Discovery 670 SPECT system equipped with a high energy general purpose (HEGP) collimator~\cite{ljungberg1989monte,morphis2021validation}. Projections were acquired at 60 angular positions spanning 360 degrees. The imaging duration for each phantom was set to 30 minutes. The photopeak window was set as 85 keV \(\pm\) 20\%, accompanied by two scattering energy windows, each 4 keV wide, positioned adjacent to either side of the photopeak energy window.
All relevant image-degrading processes were modeled, including attenuation, scatter, collimator response, septal penetration and scatter, characteristic X-ray from both the $\alpha$-emitting isotopes and the lead in the collimator, and finite energy and position resolution of the detector. The energy dependency of these image-degrading processes was also modeled. The mean of the stray-radiation-related noise in each bin, denoted as \(\psi\), was modeled as 0.18~\cite{li2022projection}. These simulation processes were validated in our previous study~\cite{li2022projection}.

\subsubsection{Figure of merit}
We evaluated the accuracy and overall error of the estimates using normalized ensemble bias (NEB) and normalized root ensemble mean squared error (NREMSE), respectively. 
Denote the estimated uptake of the $k^{th}$ region in the $c^{th}$ phantom realization, with in total $C$ considered phantom realizations, as ${\widehat{\lambda}}_{k}^{c}$, and the true value as $\lambda_{k}^{c}$. Denote ${\overline{\lambda}}_{k} = < \lambda_{k} >_{\bm{\lambda}}$. We have 
\begin{equation}
    NEB_{k} = \frac{1}{C}\sum_{c}^{C}  \frac{( {\widehat{\lambda}}_{k}^{c} - \lambda_{k}^{c} )}{{\overline{\lambda}}_{k}} 
\end{equation}
and 
\begin{equation}
    NREMSE_{k} = \frac{1}{{\overline{\lambda}}_{k}}\sqrt{\frac{1}{C}\sum_{c}^{C}( {\widehat{\lambda}}_{k}^{c} - \lambda_{k}^{c} )^{2}}.
\end{equation}

\subsubsection{Evaluating the WIN-PDQ method}

We evaluated the performance of the WIN-PDQ method by comparing its performance in estimating the regional uptake against the OSEM-reconstruction-based method~\cite{hudson1994accelerated} and the LC-QSPECT method~\cite{li2022projection}. 

\section{Results}

\begin{figure}
    \centering
    \includegraphics[width=0.85\textwidth]{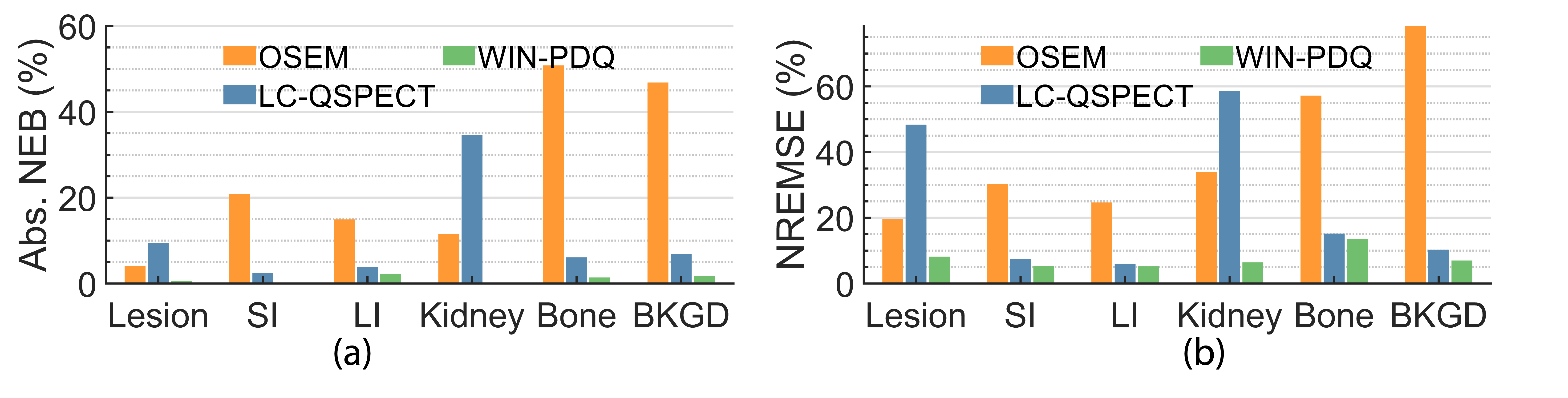}
    \caption{The (a) absolute NEB and (b) NREMSE of estimated regional uptake across all phantoms realizations, yielded by the considered quantitative SPECT methods. LI: large intestine, SI: small intestine, and BKGD: background.}
    \label{fig:std_pat}
\end{figure}

Fig.~\ref{fig:std_pat} presents the absolute NEB and NREMSE of the WIN-PDQ method in estimating the regional uptake across all phantom realizations, along with comparisons with other quantitative SPECT methods. The WIN-PDQ method demonstrates ensemble unbiasedness and consistently outperforms other quantitative SPECT methods considering the NREMSE values across all regions, especially for the lesion and kidney.

\section{Discussion and conclusion}
We have proposed the WIN-PDQ method, an approach designed to directly quantify the mean regional activity uptake in patients from SPECT projections while marginalizing intra-regional uptake heterogeneity. This method was evaluated using 3D Monte Carlo-simulated SPECT of anthropomorphic phantoms with $^{223}$Ra uptake and lumpy-model-based intra-regional uptake heterogeneity, simulating patients undergoing treatment with \(^{223}\)Ra-based \(\alpha\)-RPT.

Our preliminary results (Fig.~\ref{fig:std_pat}) show the effectiveness of the WIN-PDQ method in reliably estimating mean regional uptake in \(\alpha\)-RPT SPECT, in the presence of intra-regional uptake heterogeneity. The proposed method significantly outperformed all other considered quantitative SPECT methods. 
The WIN-PDQ method yielded nearly ensemble unbiased estimates (Fig.~\ref{fig:std_pat}a), aligning with the designs of the method. Notably, among all the considered quantitative SPECT methods, the WIN-PDQ method consistently yielded the lowest NREMSE across all regions (Fig.~\ref{fig:std_pat}b). Compared with the LC-QSPECT, the WIN-PDQ method yielded significantly lower NREMSE values in critical regions such as the lesion and the kidney. This finding is significant, considering the importance of accurate and precise lesion uptake quantification in therapy efficacy prediction and the kidney as a vital organ.
However, it was observed that the improvement in NREMSE values by the WIN-PDQ method was less significant in larger regions, such as the small and large intestines and bone, compared to LC-QSPECT. These observations, which warrant further investigation, suggest that quantification in smaller regions may be more sensitive to model mismatches related to intra-regional uptake heterogeneity. 



The proposed WIN-PDQ method requires prior knowledge of the intra-regional uptake heterogeneity model and the distribution of parameters that determine the intra-regional uptake heterogeneity (Eqs.~\eqref{eq:k_n} and~\eqref{eq:k_g_bar}). Previously developed statistical methods~\cite{kupinski2003experimental} and deep learning-based approaches~\cite{zhou2022learning} could potentially be adapted to derive the required knowledge of intra-regional uptake heterogeneity from patient population data.

With more comprehensive evaluations and validations, the proposed method could serve as an important tool for optimizing SPECT system designs. This can be achieved by modeling the impact of various system designs on the system matrix in Eq.~\eqref{eq:EMSE_ana} and identifying the system design that can yield the lowest EMSE value. A previous study used the conventional Wiener estimator for a similar purpose~\cite{lin2015using}, while the proposed WIN-PDQ method offers the opportunity to consider intra-regional uptake heterogeneity in such SPECT system optimization processes.

A limitation of the evaluation in this study is its proof-of-concept nature. The patient phantoms used were the same in body size and had identical lesion location and diameter. The promising results of this study motivate more comprehensive evaluations of the proposed method, such as through a virtual imaging trial. Furthermore, the parameters employed to simulate intra-regional uptake heterogeneity were not derived from clinical data. Future research would benefit from integrating more clinically realistic modeling of the heterogeneity.
Additionally, while the lumpy model has proven effective in modeling clinically realistic intra-regional uptake heterogeneity~\cite{liu2021observer, liu2023observer}, it is also important to extend the WIN-PDQ method to account for other models of intra-regional uptake heterogeneity. Such expansions would enhance the applicability of the method in a wider range of clinical scenarios.

To conclude, we proposed and evaluated Wiener INtegration Projection-Domain Quantification (WIN-PDQ), a Wiener-estimator-based quantitative SPECT approach for $\alpha$-particle-emitting radiopharmaceutical therapies ($\alpha$-RPTs) that directly estimates mean regional uptake from SPECT projections, accounting for intra-regional uptake heterogeneity. 
Our preliminary evaluations using three-dimensional anthropomorphic phantoms, which incorporated Gaussian-lump-based intra-regional heterogeneity, provide evidence that WIN-PDQ is ensemble unbiased and significantly outperforms the OSEM-reconstruction-based method and the LC-QSPECT method in terms of both bias and normalized root ensemble mean squared error. These promising results indicate the potential of WIN-PDQ in estimating mean regional uptake in $\alpha$-RPTs given significant intra-regional uptake heterogeneity, motivating further development and evaluation of the method, as well as application of the method to objectively optimize SPECT system and protocol designs for the task of mean regional uptake quantification.

\section*{Acknowledgements}
This work was supported by grants R01-EB031962 and R01-EB031051 awarded by the National Institute of Biomedical Imaging and Bioengineering and the NSF CAREER 2239707.





\bibliography{report} 
\bibliographystyle{spiebib} 

\end{document}